\documentclass{Interspeech}
\RequirePackage{multirow}

\interspeechcameraready

\title{Using Neurogram Similarity Index Measure (NSIM) to Model Hearing Loss and Cochlear Neural Degeneration}

\author[affiliation={1,2,3}]{Ahsan}{Cheema}
\author[affiliation={1,2,3}]{Sunil}{Puria}

\affiliation{Speech and Hearing Bioscience and Technology Program}{Harvard University}{Cambridge, USA}
\affiliation{Eaton Peabody Laboratories}{Massachusetts Eye and Ear Infirmary (MEEI)}{Boston, USA}
\affiliation{Department of Otolaryngology}{Harvard Medical School}{Boston, USA}
\email{ahsancheema@g.harvard.edu, sunil\_puria@meei.harvard.edu}

\usepackage{comment}

\begin{document}

\maketitle
\begin{abstract}
    Trouble hearing in noisy situations remains a common complaint for both individuals with hearing loss and individuals with normal hearing. This is hypothesized to arise due to condition called: cochlear neural degeneration (CND) which can also result in significant variabilities in hearing aids outcomes. This paper uses computational models of auditory periphery to simulate various hearing tasks. We present an objective method to quantify hearing loss and CND by comparing auditory nerve fiber responses using a Neurogram Similarity Index Measure (NSIM). Specifically study 1, shows that NSIM can be used to map performance of individuals with hearing loss on phoneme recognition task with reasonable accuracy. In the study 2, we show that NSIM is a sensitive measure that can also be used to capture the deficits resulting from CND and can be a candidate for noninvasive biomarker of auditory synaptopathy.
\end{abstract}
\section{Introduction}
Difficulty understanding speech in the presence of background noise is one of the most common complaints of patients with sensorineural hearing loss (SNHL) \cite{Vermiglio2012-fi}. Extensive research has shown that the death of cochlear hair cells leads to hearing loss, but it is often preceded by a loss of the synapses linking the hair cells to the auditory-nerve fibers (cochlear neural degeneration). Cochlear neural degeneration (CND) can contribute to difficulty hearing in noise and goes undetected during standard audiometric evaluations  \cite{Kujawa2009-nv}. Hearing aids (HAs) remain the standard treatment for hearing loss, but they primarily address hearing loss due to death of cochlear cells, leaving neural deficits untreated. This limits the effectiveness of hearing aids and can result in variability in patient outcomes. Therefore, development of an objective measure to study the effects of hearing loss and CND remains an essential challenge which can help improve the design and gain compensation strategies for hearing aids.\\
One of the promising approaches to arrive at an objective measures for hearing loss and CND is to simulate the response of auditory nerve fibers (ANF) from a normal cochlea and a cochlea with hearing loss using physiologically inspired auditory nerve model (ANM) of cochlea \cite{Bruce2018-iv} and compare differences between the discharge patterns for the two cases. In order to study the differences in discharge patterns of ANF Hines and Harte \cite{Hines2010-cn, Hines2012-nj} developed a "Neurogram Similarity Index Measure (NSIM)" which is similar to Structural Similarity Index Measure (SSIM) used widely in the image processing literature \cite{Wang2004-uo}. This measure views the time-frequency discharge patterns from auditory nerve fibers as an image pattern called a ‘neurogram’ and can be used to quantify the differences between ANFs from a normal hearing cochlea with no hearing loss and a cochlea with hearing loss. Previous studies have established NSIM as a quantitative measure to simulate performance of normal hearing individuals on /CVC/ phoneme recognition tasks for a range input stimulus levels \cite{Hines2012-nj}. Mamun et. al. \cite{7035038} improved the dynamic range for neurogram based measures using Neurogram orthogonal polynomial to map performance of normal hearing individuals, but this study only included 1 hearing loss individual and thus was not generalizable to study the effect of hearing loss. No previous study has comprehensively (on a large enough datasets of individuals with hearing loss) explored the utility of neurogram based objective measures to study the effects of hearing loss on ANF responses and link that to the performance on speech recognition tasks. Similarly, the effects of CND on neurogram based metrics has also not been explored in previous literature. In this work we present two studies that evaluate NSIM as a metric. In Study 1, NSIM is used to evaluate performance of individuals with varying degrees of hearing loss. In Study 2, we use NSIM to model varying degrees of CND. 
\section{Methods}
\subsection{Phenomenological Model of Cochlea and Auditory Nerve Fiber Response}
This study used ANM \cite{Bruce2018-iv} to simulate the responses of ANF and  generate neurogram for a given sound (.wav file) and for a specific hearing loss profile. The model allows the simulation of hearing loss by decreasing OHC gain and making the frequency response of the cochlear filters broader \cite{Bruce2018-iv}. The output of the ANM is in the form of post stimulus histogram (PSTH) which represents the spiking activity of the auditory nerve fibers in time bins. Each time bin in the PSTH can be converted to frequency response and generate a neurogram, which maps the spiking activity on a time-frequency map. The ANM allowed simulation of the three different types of auditory nerve fibers: 1) Low Spontaneous (LS) rate fibers with high thresholds, 2) Medium Spontaneous (MS) rate Fibers with medium thresholds, and 3) High Spontaneous (HS) rate fibers with low thresholds. The summed spiking activity of all fibers can be used to get one neurogram that represents activity for all three ANF types in response to an acoustic stimulus, or three separate neurograms representing each ANF fiber types. Additionally, to capture long term and short-term properties of auditory nerve, the model allows control of the time bins used for the PSTH which in turn allows generation of fine timing (FT) neurograms and mean rate (MR) neurograms. 
\subsection{Neurogram Similarity Index Measure:}
\begin{figure}[htbp]
    \centering
    \includegraphics[width=0.8\linewidth]{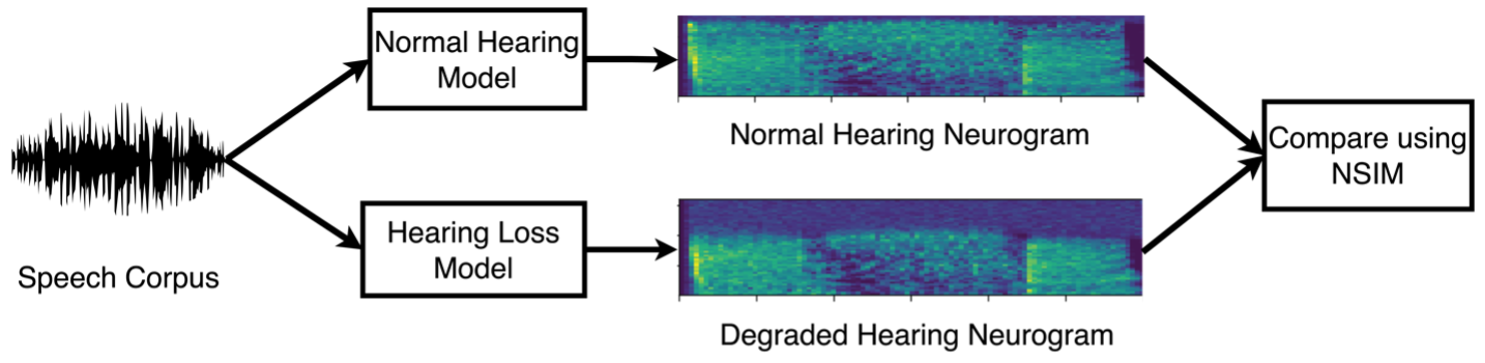} 
    \caption{Design of experiment to simulate Neurogram Similarity Index Measure}
    \label{fig:nsim_experiment}
\end{figure}

To quantify the similarity between the normal hearing neurograms and degraded hearing neurograms (neurograms from a hearing loss cochlea), methods previously developed \cite{Wang2004-uo, Hines2010-cn} were used to calculated NSIM. The image processing based method Structural Similarity Index Measure (SSIM) compares luminance(l), contrast(c) and structure(s) between a reference image and degraded image using a suitable window size \cite{Wang2004-uo} using Eq. \ref{eq:ssi}. The constants $\alpha$, $\beta$, and $\gamma$ are the weighting coefficients that are applied to weigh the different components of the similarity equation. The coefficients $C_{1}, C_{2}, C_{3}$ have negligible effect on the results and are added to prevent unstable results at boundaries. Hines et. al \cite{Hines2012-nj} applied the same method and derived an equivalent similarity measure for comparing neurograms. They showed that for the neurograms, in the Eq. \ref{eq:ssi} contrast and component weightings have negligible effect. This reduces Eq. \ref{eq:ssi} to Eq. \ref{eq:nsi} which is then calculated over a 3X3 gaussian window of radius 0.5 \cite{Hines2012-nj, 7940042} and averaged over all the points in the neurogram to calculate the NSIM (eq. \ref{eq:nsim}). The general design of experiments is given in Fig. \ref{fig:nsim_experiment}.  
\begin{equation}
SSI(r,d) = \left[l(r,d)\right]^\alpha \cdot \left[c(r,d)\right]^\beta \cdot \left[s(r,d)\right]^\gamma
\label{eq:ssi}
\end{equation}
where: 
\[
l(r,d) = \frac{2\mu_r \mu_d + C_1}{\mu_r^2 + \mu_d^2 + C_1} , \\ 
c(r,d) = \frac{2\sigma_{rd} + C_2}{\sigma_r^2 + \sigma_d^2 + C_2} , \\
\]
\[s(r,d) = \frac{\sigma_{rd} + C_3}{\sigma_r \sigma_d + C_3}, \\\]
\[
\mu_r = \text{mean of reference image intensity over window},
\]
\[
\mu_d = \text{mean of degraded image intensity over window},
\]
\[
\sigma_r = \text{standard deviation of reference image over window},
\]
\[
\sigma_r^2 = \text{variance of reference image over window},
\]
\[
\sigma_d = \text{standard deviation of degraded image over window},
\]
\[
\sigma_d^2 = \text{variance of degraded image over window},
\]
\[
\sigma_{rd} = \text{covariance matrix over the window},
\]
\[
C_{1} = 0.01L, \ \ \ C_{2} = (0.03L)^{2}, \ \ \ C_{3} = \dfrac{C_{2}}{2}, \ \ \
\]
\[ L = intensity \ range \]
\begin{equation}
NSI(r,d) = \frac{2\mu_r \mu_d + C_1}{\mu_r^2 + \mu_d^2 + C_1} \cdot \frac{\sigma_{rd} + C_3}{\sigma_r \sigma_d + C_3}
\label{eq:nsi}
\end{equation}
\begin{equation}
\text{NSIM}(r, d) = \frac{1}{N \cdot M} \sum_{f=1}^{N} \sum_{t=1}^{M} \text{NSI}(r, d)
\label{eq:nsim}
\end{equation}
\[ N = total \ frequency \ bins ,\\\ M = total \ time \ bins \]

\subsection{Study 1: Correlating phoneme-recognition-task performance with NSIM}
For this study, speech material selected and tested by Harris et. al. \cite{Hajicek2023-oh, harris2024evaluation} was used. The speech level was administered monoaurally at 65 dB SPL.  The test consisted of presenting /VCV/ speech stimuli with the vowel /a/ and 10 different consonants combined with speech shaped noise at different signal to ratios (SNR) found to be the most sensitive to predict hearing loss \cite{Hajicek2023-oh}. In their study there were a total of 94 individuals with hearing loss ranging between Normal hearing ( PTA = $\leq$ 15 dB HL) to Moderate Hearing loss (PTA = 41 -- 55 dB HL). Corpus developed by Harris et. al. \cite{harris2024evaluation} was used here and processed through ANM to simulate 94 hearing loss profiles that were clinically measured by Harris et. al. Neurograms were generated for each hearing loss profile and compared with corresponding normal hearing neurogram using NSIM. For each hearing loss profile/individual, MR-NSIM and FT-NSIM were calculated for all 10 /VCV/ from Harris et. al. and averaged to generate average MR-NSIM and FT-NSIM for each hearing loss profile/individual. \\
Average MR and FT NSIM values across all 10 phonemes and Pure Tone Average Thresholds (PTA in dB) were used to train support vector regression(SVR) based models to predict performance of individuals as recorded by Harris et. al. \cite{harris2024evaluation} using a threefold cross validation paradigm (Fig. \ref{fig:nsim_model_performance}).

\subsection{Study 2: Extending NSIM to study Cochlear Neural Degeneration (CND)} CND is characterized by loss of auditory nerve synapses which ultimately results in less information reaching the higher order processing centers in the brain. To simulate the performance that can be used to predict and study CND, it is important to select a type of stimulus that can encode the effects of CND at the neurogram level. Additionally, all three types of fibers have different response characteristics and are present in unequal numbers in an AN bundle. Therefore, to accurately study the effect of CND, it may be important to look at neurograms for each fiber type independently and then compare them with the corresponding fiber type neurogram from a normal hearing cochlea. The resulting similarity maps for each fiber type can then be summed and averaged to get one value of similarity for the ease of analysis. Another hypothesis related to effects of CND is the decrease in information content in the neural signal which becomes more evident only during complex listening tasks \cite{DiNino2022-so}. Therefore, in order to simulate these conditions, 20 lists consisting of 10 iso-phonemic, monosyllabic common /CVC/ words were used. There were no word repetitions across the lists. These words were generated using Google text-to-speech api in python \cite{gtts}. After generating the wav files for the words, they were normalized using the RMS value of the waveforms (16 bit) and converted to pressure values using:
\begin{equation}
\mathrm{p} = \frac{\mathrm{wav}}{\mathrm{rms}(\mathrm{wav})} \times 20 \times 10^{-6} \times 10^{\frac{L}{20}}
\label{eq:stim_scaling}
\end{equation}
where $L \in \{50,\,65,\,80,\,95\}~\mathrm{dB\,SPL}$.

These words were presented in a range of listening levels from 50—95 dB SPL, and difficulty conditions: 1) without any compression or reverberation, 2) with 65\% time compression, and 3) with 65\% time compression and reverberation. Audiometric profile shown in Table \ref{tab:audiometric profile A} and different degrees of CND as shown in the Table \ref{tab:CND_profiles} were used in this study. The NSIM for each fiber type were calculated separately for all 200 words and then averaged to get a single NSIM values for FT-neurogram and MR-neurograms for each fiber type (eq. \ref{eq:overall_nsim}). As a next step, the NSIM values for the sloping loss profile with no CND were used as baseline values and then the effect of introducing CND was quantified by normalizing with CND to hearing loss profile without any CND (eq. \ref{eq:cnd_effect}).
\begin{equation}
NSIM_{\mathrm{overall}} = \frac{1}{3} \left( NSIM_{\mathrm{LS}} + NSIM_{\mathrm{MS}} + NSIM_{\mathrm{HS}} \right)
\label{eq:overall_nsim}
\end{equation}
\begin{equation}
CND_{\mathrm{effect}} = \dfrac{NSIM_{HL \ no\  CND} - NSIM_{HL \ and \ CND}}{NSIM_{HL \ no\  CND}}
\label{eq:cnd_effect}
\end{equation}
\begin{table}[ht]
\centering
\caption{Hearing Loss profile used in the study and simulated using OHC loss in the cochlear model}
\label{tab:audiometric profile A}
\scalebox{0.6}{%
\begin{tabular}{lccccccc}
\toprule
Audiometric Profile & 125 Hz & 250 Hz & 500 Hz & 1000 Hz & 2000 Hz & 4000 Hz & 8000 Hz \\
\midrule
Sloping Loss        & 0      & 0      & 10     & 20      & 23      & 45      & 75      \\
\bottomrule
\end{tabular}%
}
\end{table}

\begin{table}[ht]
\centering
\caption{CND Profiles used for the age-related hearing loss profile listed in Table 1. The distribution [Low-SR, Med-SR, High-SR] = [5, 5, 12] refers to the number of low-spontaneous-rate (LS), medium-spontaneous-rate (MS), and high-spontaneous-rate (HS) fibers, respectively, for the normal cochlea with no CND.}
\label{tab:CND_profiles}
\scalebox{0.55}{%
\begin{tabular}{llccc}
\toprule
Audiometric Profile & CND Profile                     & Low-SR Fibers & Med-SR Fibers & High-SR Fibers \\
\midrule
\multirow{7}{*}{Sloping Loss} & No CND                         & 5             & 5             & 12           \\
                              & 20\% LS MS loss                & 4             & 4             & 12           \\
                              & 40\% LS MS loss                & 3             & 3             & 12           \\
                              & 60\% LS MS loss                & 2             & 2             & 12           \\
                              & 80\% LS MS loss                & 1             & 1             & 12           \\
                              & 100\% LS MS loss               & 0             & 0             & 12           \\
                              & 100\% LS MS loss, 20\% HS loss & 0             & 0             & 10           \\
\bottomrule
\end{tabular}%
}
\end{table}

\section{Results}
\subsection{Study 1: Correlating phoneme-recognition-task performance with NSIM}
\begin{figure}[htbp]
    \centering
    \includegraphics[width=1.05 \linewidth]{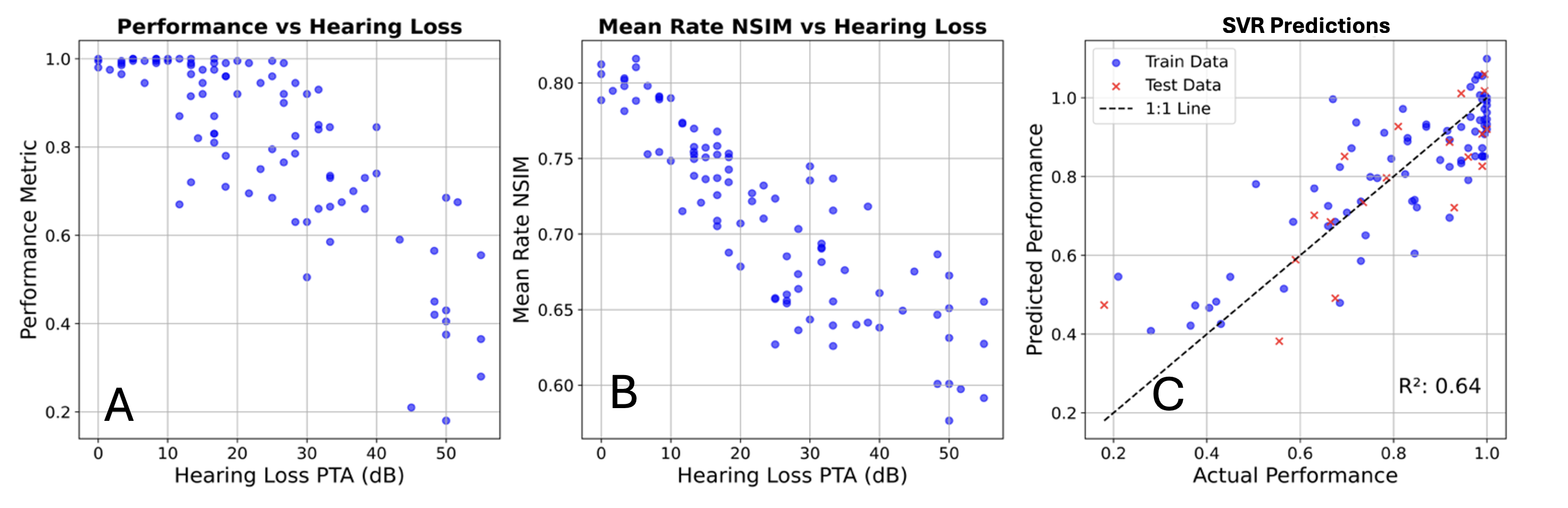} 
    \caption{(A) Percent-correct performance measured on 94 subjects. (B) Mean Rate NSIM on the phoneme-recognition task plotted against the Pure Tone Average (PTA) threshold hearing loss (dB) for 94 subjects. (C) Performance of the support-vector regression (SVR) predictions trained using the Mean Rate NSIM, Fine Timing NSIM, and PTA as features resulting in R2=0.64.}
    \label{fig:nsim_model_performance}
\end{figure}

We trained support vector regression-based models (Fig. \ref{tab:svr_hyperparams}) to predict performance on a phoneme recognition task using the MR-NSIM, FT-NSIM and PTA (dB) values (Table \ref{tab:svr_hyperparams}). The results from the best performing model are presented in the Fig. \ref{fig:nsim_model_performance}C. The model presented achieved a mean squared error of 0.015 and $R^2$ value of 0.64. The models which incorporated both MR-NSIM and FT-NSIM features performed the better as shown in the Table \ref{tab:svr_hyperparams}. There is a discrepancy between the variation of MR-NSIM and the actual performance when plotted against pure-tone average hearing loss (dB) (Fig. \ref{fig:nsim_model_performance}B), particularly below 25 dB HL. MR-NSIM degrades monotonically as hearing loss increases, whereas actual performance only begins to decline beyond approximately 25 dB HL. We attribute this difference to the way NSIM is calculated, which uses the ideal (no-loss) neurogram as the reference (eq. \ref{eq:nsim}) and compares it to degraded neurograms. To address this, we do not use raw NSIM values directly as a measure of performance; instead, we treat them as features in our downstream SVR models (Table \ref{tab:svr_hyperparams}). 
\begin{table}[ht]
\centering
\caption{Performance of best Performing Support Vector Regression Models trained on various feature combinations to predict performance on phoneme recognition task}
\label{tab:svr_hyperparams}
\scalebox{0.55}{%
\begin{tabular}{l l c c}
\toprule
\textbf{Features} & \textbf{Hyperparameters} & \textbf{MSE} & \textbf{R\textsuperscript{2}} \\
\midrule
MR NSIM & \{\,C: 1, \,epsilon: 0.075, \,gamma: auto, \,kernel: rbf\} & 0.023 & 0.485 \\
FT NSIM & \{\,C: 0.25, \,epsilon: 0.1, \,gamma: auto, \,kernel: rbf\} & 0.028 & 0.352 \\
MR NSIM $\times$ FT NSIM & \{\,C: 10, \,epsilon: 0.05, \,gamma: auto, \,kernel: rbf\} & 0.018 & 0.588 \\
MR NSIM $\times$ FT NSIM $\times$ PTA & \{\,C: 2.5, \,epsilon: 0.05, \,gamma: scale, \,kernel: linear\} & 0.0157 & 0.639 \\
\bottomrule
\end{tabular}%
}
\end{table}
\subsection{Study 2: Extending NSIM to study Cochlear Neural Degeneration}
\begin{figure}[htbp]
    \centering
    \includegraphics[width=1\linewidth]{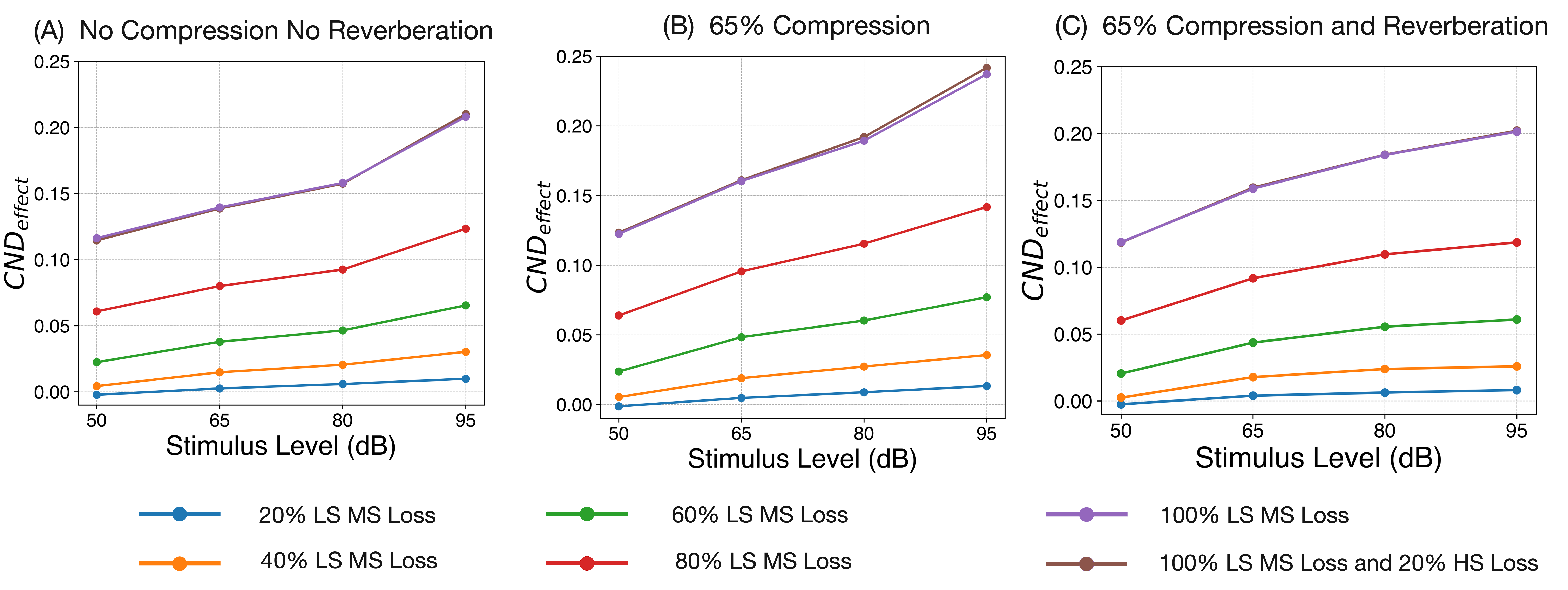} 
    \caption{Average $CND_{effect}$ calculated from MR-NSIM for the entire speech corpus (indicated in the title of subplots) and plotted across various signal levels (dB SPL) for Hearing Loss profiles with varying degrees of CND as indicated by the legend.}
    \label{fig:CND_effect}
\end{figure}

Figure \ref{fig:CND_effect} shows the CND$_{effect}$ using Eq. \ref{eq:cnd_effect} (with degree of fiber loss defined in Table \ref{tab:CND_profiles}) plotted against the signal level (dB SPL), for the three different speech types in each panel. For every step increase in CND by 20\%, there is a change in the values the CND$_{effect}$ (different colored lines) at all the stimulus levels tested and all three speech types of speech material. The greatest effect due to CND was observed at 95 dB SPL for all speech materials. The overall effect of CND is highest for the 65\% compressed speech for profiles with total loss of LS and MS fibers (light purple). Additional loss of 20\% HS fibers did not affect the results significantly (near overlap between light and darker purple lines).

\section{Discussion}
Zaar and Carney \cite{Zaar2022-sd} predicted speech-reception thresholds (SRTs) for subjects with normal hearing or hearing loss using different types of speech material and speech-in-noise conditions using neurograms fed into an approximate model for the inferior colliculus (IC). Our strategy differs from \cite{Zaar2022-sd} in that our backend decision is based on NSIM rather than correlation on approximate responses of IC. As such, our approach only uses responses from the auditory periphery, which is relatively well understood, and is free from the potential drawbacks of assumptions and hypotheses about processing strategies used by central auditory brain mechanisms. Another important difference from previous work is that the combined ANM and NSIM methods have not been applied to quantify the effects and contributions of CND.
\subsection{Study 1: Correlating phoneme-recognition-task performance with NSIM} In Study 1 we were able to show that SVR models trained using both the MR-NSIM, which represents the long-term similarity in the response, and FT-NSIM, which represents short term similarity between neurograms, performed the better when predicting performance on the phoneme recognition task. This shows that for a phoneme recognition task, individuals rely on both short-term spiking differences to detect phoneme transitions particularly between vowel and consonant and long term spiking activity to track the low-frequency components in the speech. The R$^{2}$ value for the best trained model was ~0.64. The performance of the model was worst for region where there were fewest points in the training data (Fig \ref{fig:nsim_model_performance}). The results from this study show that NSIM is a useful measure to study and simulate performance of individuals with hearing loss and given enough data, it can generate reliable predictions for performance on phoneme recognition task. However, one caveat remains: we observed that for lower degrees of hearing loss, the MR NSIM values degrade but the actual performance does not degrade (Fig. \ref{fig:nsim_model_performance}A,B) which suggests that there could be certain thresholds for the raw NSIM values which must be exceeded before the performance begins to degrade.

\subsection{Study 2: Extending NSIM to study Cochlear Neural Degeneration (CND)} In Study 2, we demonstrated that when using speech recognition tests to detect the effect of CND or synaptopathy, the highest differences between profiles with and without CND were observed when the presentation level of the stimulus was high (above 80 dB SPL), which follows from our understanding of that at higher stimulus levels, all three types of fibers are required to encode the information, since the HS fibers with low thresholds will be saturated and to increase the total number of spikes, LS/MS fibers will need to be recruited. However, in the absence of these fibers (typical CND case) no additional spiking activity is produced resulting in lower NSIM and hence greater difference between profiles with and without CND. It was also observed that the maximum effect of CND was observed for 65\% compressed speech without reverberation, which means that 65\% compressed speech test can be used to discriminate CND profile from the profile without CND. Increasing the difficulty of test further by adding reverberation slightly decreases the CND$_{effect}$. This means that even though the difficulty of the test will increase by adding reverberation but without any CND discriminability benefits. This would mean more false positives, if such a test were used as a metric to detect CND in patients. \\
Another interesting observation from the Fig. \ref{fig:CND_effect} is that for all speech material, the values of CND$_{effect}$ are relatively small up to 60\% LS MS Loss profile (green line), however 80\% CND profile has approximately twice the values for CND$_{effect}$ at all levels for all profiles. This observation is similar to what Grantt et. al. \cite{Grant2022-zs} observed where they saw significant reduction in word recognition scores only after the individual had lost more than 60\% neurons. This further suggests that NSIM is a reliable measure and can be employed as a sensitive metric for detecting CND.
\section{Conclusions and Future Directions:}
Both the studies indicate that NSIM can be used as an objective metric to map performance on phoneme recognition tasks and to detect sensory hearing loss associated outer hair cell loss (Study 1) and it can be used to estimate CND (study 2). In study 1, we observed that for lower degrees of hearing loss, there is difference in the trend of actual performance and raw NSIM values (Fig. \ref{fig:nsim_model_performance} A,B). Future work can explore deriving thresholds that can be applied on the raw NSIM values, so that it can be related to the performance on speech tests. In the study 2, only 1 hearing loss profile was simulated, and future studies can use the methods presented here to look at all commonly observed hearing loss profiles. Additionally, average values for NSIM were calculated by averaging NSIM for all fiber types, however, there is no physiological evidence that differences/correlations for all three types of fibers can be averaged. In this study average NSIM for all three fiber types simplified the analysis. Future studies can explore deriving weighting coefficients for the NSIM for different fiber types. Another possible future direction would be to conduct the speech tests on human subjects with material discussed in study 2 and relate the NSIM to performance measures on these tests. Having established NSIM as an objective measure can ultimately allow deriving hearing aids gain compensation strategy that can account for CND, which currently remains a major challenge in hearing healthcare.

\section{Acknowledgements}
The authors would like to thank Dr. Stephen Neely from Boystown National Research Hospital for sharing the dataset used for this study. We would also like to acknowledge our funding sources: National Institutes of Health (NIH), National Institute on Deafness and Other Communication Disorders grant number R01DC007910-16, Massachusetts Eye and Ear Infirmary (MEEI) Shark Tank Research Award, and funding from Harvard Graduate School of Arts and Sciences (GSAS). 


\bibliographystyle{IEEEtran}
\bibliography{output}

\end{document}